\documentclass[sn-mathphys,Numbered]{sn-jnl}

\usepackage{graphicx}%
\usepackage{multirow}%
\usepackage{amsmath,amssymb,amsfonts}%
\usepackage{amsthm}%
\usepackage{mathrsfs}%
\usepackage[title]{appendix}%
\usepackage{xcolor}%
\usepackage{textcomp}%
\usepackage{manyfoot}%
\usepackage{booktabs}%
\usepackage{algorithm}%
\usepackage{algorithmicx}%
\usepackage{algpseudocode}%
\usepackage{listings}%
\usepackage{lineno}

\theoremstyle{thmstyleone}%
\theoremstyle{thmstyletwo}%

\theoremstyle{thmstylethree}%

\begin{document}
\linespread{1.1}
\title[Thin film aluminum nitride surface acoustic wave resonators for quantum acoustodynamics]{Thin film aluminum nitride surface acoustic wave resonators for quantum acoustodynamics}


\author[1,2]{\fnm{Wenbing} \sur{Jiang}}
\equalcont{These authors contributed equally to this work.}
\author[1,2]{\fnm{Junfeng} \sur{Chen}}
\equalcont{These authors contributed equally to this work.}

\author[1,2]{\fnm{Xiaoyu} \sur{Liu}}

\author[1,3]{\fnm{Zhengqi} \sur{Niu}}

\author[1,2]{\fnm{Kuang} \sur{Liu}}

\author[1,2]{\fnm{Wei} \sur{Peng}}

\author[1,2]{\fnm{Zhen} \sur{Wang}}

\author*[1,2]{\fnm{Zhi-Rong} \sur{Lin}} \email{zrlin@mail.sim.ac.cn}

\affil[1]{\orgdiv{State Key Laboratory of Functional Materials for Informatics}, \orgname{Shanghai Institute of Microsystem and Information Technology, Chinese Academy of Sciences}, \orgaddress{\city{Shanghai}, \postcode{200050}, \country{China}}}

\affil[2]{\orgname{University of Chinese Academy of Science}, \orgaddress{\city{Beijing}, \postcode{100049}, \country{China}}}

\affil[3]{\orgname{ShanghaiTech University}, \orgaddress{\city{Shanghai}, \postcode{201210}, \country{China}}}


\abstract{The quantum excitations of macroscopic surface acoustic waves (SAWs) have been tailored to control, communicate and transduce stationary and flying quantum states. However, the limited lifetime of this hybrid quantum systems remains critical obstacles to extend their applications in quantum information processing. Here we present the potentials of thin film aluminum nitride to on-chip integrate phonons with superconducting qubits over previous bulk piezoelectric substrates. We have reported high-quality thin film GHz-SAW resonators with the highest internal quality factor $Q_\textrm{i}$ of $5 \times 10^{4}$ at the single-phonon level. The internal losses of SAW resonators are systematically investigated with tuning the parameters of sample layout, power and temperature. Our results manifest that SAWs on piezoelectric films are readily integrated with standard fabrication of Josephson junction quantum circuits, and offer excellent acoustic platforms for the high-coherence quantum acoustodynamics architectures.}




\maketitle

\section{Introduction}\label{sec1}

Surface acoustic wave (SAW) resonators in the quantum regime are of tremendous interest owing to their extensive applications in superconducting quantum processing, storage and communication in recent decades \cite{aref2016quantum,delsing2019,clerk2020hybrid,cleland2004superconducting,chu2020perspective}. The slow propagation speed and the short wavelength of SAWs compared to electromagnetic waves enable the novel atomic physics phenomena, such as nonclassical phonon states \cite{satzinger2018quantum}, routing of propagating phonons \cite{ekstrom2019towards}, electromagnetically induced acoustic transparency \cite{andersson2020electromagnetically} and non-Markovian dynamics of the giant atom \cite{andersson2019non}, etc. Of particular interest in SAWs are the concurrence of inherent piezoelectric and photoelastic effects, rendering SAWs to be promising ingredients towards interlinking the disparate quantum systems \cite{shumeiko2016quantum,okada2021superconducting}. In above cases, the strong coupling between the SAW phonon and superconducting qubits need be achieved primarily, as has been recently demonstrated in the SAW waveguide or resonator on piezoelectrical crystals coupled to the superconducting transmon qubit \cite{satzinger2018quantum,gustafsson2014propagating,manenti2017circuit,moores2018cavity,bolgar2018quantum}. However, in such quantum acoustodynamics (cQAD) systems, the quantum state lifetime is remarkably limited in contrast with that in circuit quantum electrodynamics (cQED) systems. The qubit relaxation time $T_1$ in the SAW-qubit coupling system rarely reaches 1 $\mu$s, ultimately restricting the performance of advanced quantum operations in the SAW-qubit hybrid system, e.g. synthesizing higher phonons Fock states \cite{satzinger2018quantum} and high-fidelity quantum state swap between microwave photons and phonons \cite{bienfait2019phonon,dumur2021quantum}.

So far there exist two main approaches, depending on the device construction, to realize the strong coupling between the SAW phonon and superconducting qubits. The first approach is to monolithically integrate all the circuits on bulk piezoelectric substrates, e.g., GaAs and ST-X quartz. The other utilizes the flip-chip assembly where the qubits and SAW circuits are separately fabricated on the low-loss sapphire and piezoelectric lithium niobate \cite{satzinger2018quantum,bienfait2019phonon,dumur2021quantum}. For the former, the qubit lifetime is drastically limited by the dielectric loss due to the piezoelectricity \cite{scigliuzzo2020phononic}. In the latter case, despite the fact that the qubit coherence time is effectively improved, the internal quality $Q_\textrm{i}$ of SAW resonators is still not optimized well to realize high fidelity quantum state transfers and remote entanglement. Consequently, developing new coupling schemes and decreasing the internal loss of SAWs remain vital but challenging topics \cite{manenti2016surface,magnusson2015surface}. Recently, with the optimization of phononic structures and fabrication technologies for the SAW circuit \cite{shao2019phononic}, the internal quality factor $Q_\textrm{i}$ has been improved to be order of $10^5$ with the resonance frequency $f_0$ $\sim$ 1 GHz. However, at higher frequency range exceeding 4 GHz where the superconducting transmon/Xmon qubits operate, $Q_\textrm{i}$'s of the SAW resonators are still low.

Integrating the SAW and superconducting qubits on piezoelectric thin films has been intuitively proposed to construct on-chip high-coherence superconducting cQAD devices, where the qubits could be placed on the desired non-piezoelectric region via selective etching \cite{andersson2021acoustic,luschmann2023surface}. In this letter, we experimentally report the nanofabrication and the phonon loss characterizations of high-frequency SAW resonators on piezoelectric AlN films. We employ AlN-on-sapphire as the substrate to design and fabricate SAW resonators because of the high SAW velocity ($v \simeq$ 5600 m/s) and strong piezoelectricity of AlN \cite{delsing2019,bu2006surface}, along with low dielectric loss in sapphire \cite{megrant2012planar}. Our results firmly validate the advantages of the thin film SAW to realize high-coherence and strong-coupling regimes in the monolithic integrated SAW-qubit chip. We demonstrate SAW resonators on piezoelectric AlN films with the highest single-phonon $Q_\textrm{i}$ of $5 \times 10^4$ in the quantum regime, and investigate the phonon loss mechanism of acoustic resonators in more detail, which are different from that of well-studied superconducting microwave resonators \cite{sage2011study,mcrae2020materials}.

\section{Results}\label{sec2}

Our SAW resonators are fabricated on the AlN-on-sapphire wafer from Kyma Technologies, Inc.,  with a top layer of 400 nm of c-axis AlN film deposited on the 420 $\mu$m thick c-plane sapphire substrate. The profile structure of the SAW resonator,
as depicted in Figure 1a, consists of two parts: one or two interdigital transducers (IDTs) depending on the device functionality, and two symmetrically distributed Bragg mirrors faced each other. IDTs comprised of periodical arrayed alter-polarity electrodes are the core component to excite and detect the SAW via the piezoelectric effect. Applying a high-frequency AC voltage to the IDT, the SAW can be excited and subsequently propagates along the solid surface. The faced Bragg mirrors formed by hundreds of regular array of shorted metallic stripes confine the activated SAW tightly, forming an acoustic resonator. The one-port SAW resonators on AlN-on-sapphire shown in Figure 1b are fabricated by two-step lift-off techniques (see supplementary material for the complete fabrication procedures).

\begin{figure}[htbp]
  \centering
  \vspace{0cm}
  \includegraphics[width=\linewidth]{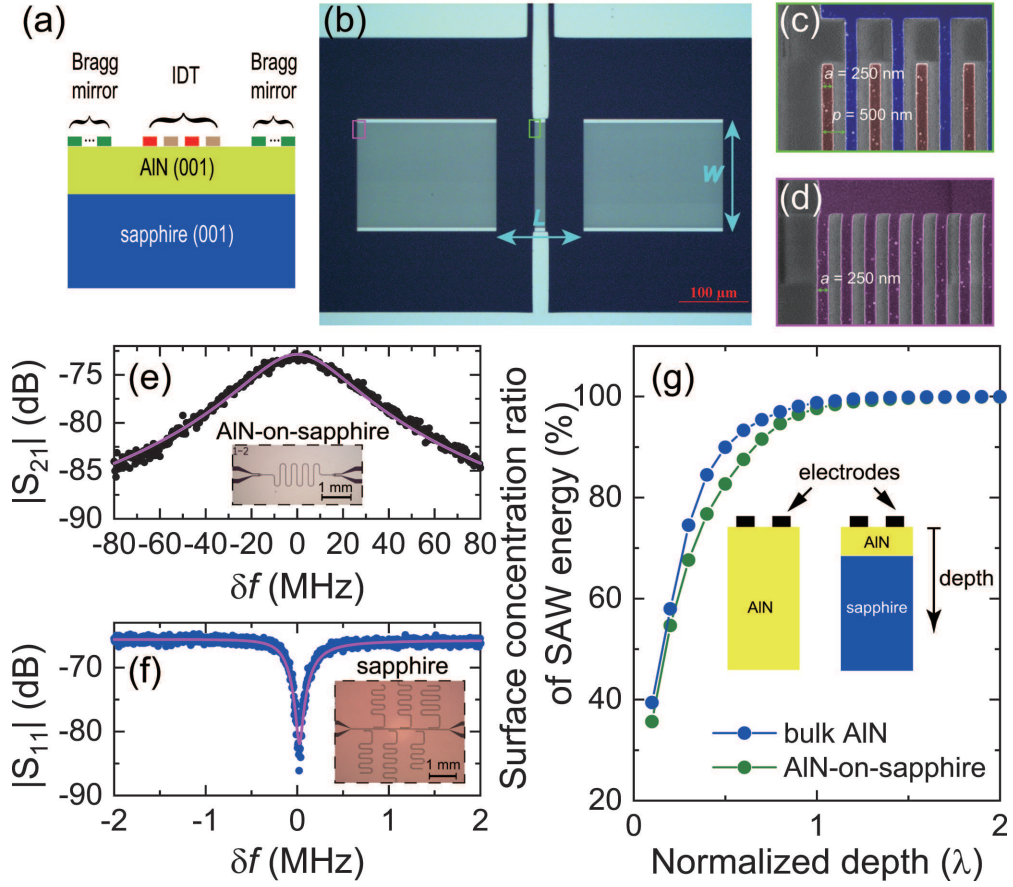}
  \caption{ AlN films SAW devices and the potentials for integrating on-chip high-coherence qubits. (a) Schematic profile of the SAW resonator on the piezoelectric AlN films. (b) False-color optical micrograph of the one-port SAW resonator fabricated on AlN-on-sapphire. (c)-(d) False-color scanning electron microscope (SEM) images of the IDT (green) and the mirror (magenta), respectively. The width of the electrode and the pitch are denoted, \emph{a} = 250 nm and \emph{p} = 500 nm, hence the SAW resonance frequency $f=\frac{v}{2p}$ is about 5.6 GHz. (e)-(f) The measured \emph{S}-parameters and the corresponding fits (magenta) for superconducting CPW resonators on the primary substrate and exposed sapphire after AlN-etching so as to quantitatively verify their dielectric loss. The insets depict optical micrographs of the fabricated CPW resonators. (g) The calculated surface concentration ratio of SAW energy with respect to the normalized depth for bulk AlN (blue) and AlN-on-sapphire (green).}
  \label{}
\end{figure}

Next, we discuss the advantages of SAW resonators on AlN films in contrast with bulk piezoelectric substrates for integrating on-chip high-coherence qubits. Piezoelectric substrates have immense microwave dielectric loss due to the electromechanical conversion from microwave photons to bulk and surface acoustic waves \cite{scigliuzzo2020phononic}. Generally, placing the qubit circuits on the low-loss dielectric substrate after selectively etching piezoelectric films would substantially increase the coherence time. Herein, we fully validate the speculation by the viable wet etching of AlN (see supplementary material) and quantifying the dielectric loss with measuring $Q_\textrm{i}$'s of the superconducting coplanar waveguide (CPW) resonators on the primary substrate and AlN-etched sapphire, respectively. Figure 1e and 1f display the measured \emph{S}-parameters and fittings, deriving their $Q_\textrm{i}$'s of 200 and $1.125 \times 10^5$ at single-photon levels, respectively. The improvement of $Q_\textrm{i}$ by three orders of magnitude in AlN-etched sapphire envision improving the qubit $T_1$ in the SAW-qubit coupling experiments by patterned etching of AlN films. Furthermore, we calculated the surface concentration ratios of SAW energy for AlN-on-sapphire and bulk AlN to compare their intrinsic phonon loss, as shown in Figure 1g. SAW energy in AlN confines more in the surface from the results, which means the internal \emph{Q}-factor of SAW resonators in AlN should be higher than that in AlN-on-sapphire. However, we note that the surface concentration ratios of SAW energy in AlN-on-sapphire is just slightly smaller than that in bulk AlN, indicating thin film structures of AlN-on-sapphire will not greatly deteriorate the internal \emph{Q}-factor of SAW modes.

\begin{figure}[!htbp]
  \centering
  \vspace{0cm}
  \setlength{\belowdisplayskip}{0cm}
  \includegraphics[width=\linewidth]{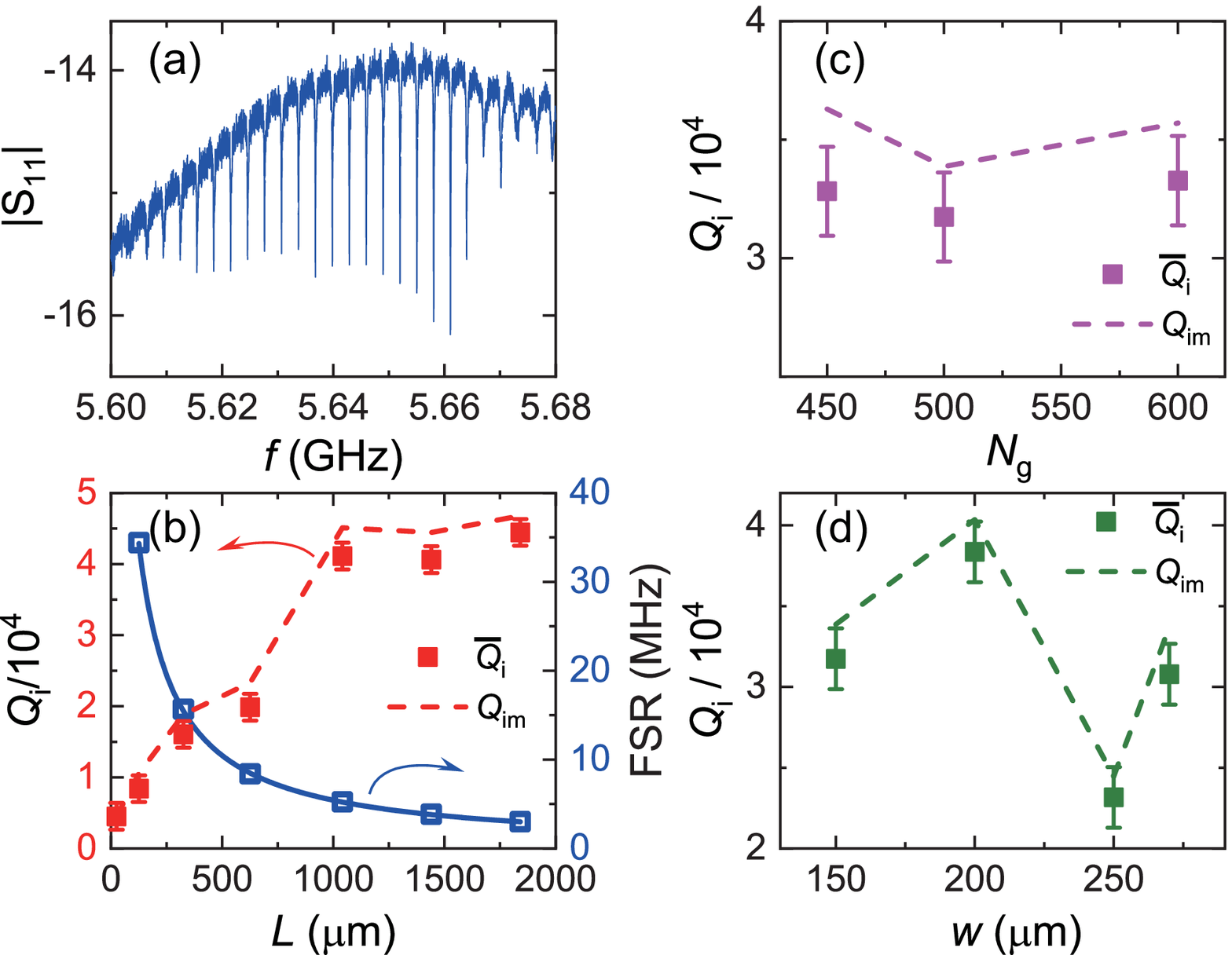}
  \caption{ Measured $Q_\textrm{i}$ of SAW resonators and the optimization with the geometry layout. (a) Magnitude of the measured reflection coefficient $|S_\textrm{11}(f)|$ of the device S1 with the input power at the single-phonon level. (b) The extracted $\overline{Q_\textrm{i}}$ and the FSR as a function of the SAW cavity length \emph{L}. The measured devices are picked from the same wafer W1. The blue solid line is an inverse proportional fit to $\delta f/f_0 =1/(L/p+1/|r_\textrm{s}|)$. The mean $\overline{Q_\textrm{i}}$ for multi-modes of SAW resonators and the maximum value $Q_\textrm{im}$ versus (c) the periods of each reflective mirror $N_\textrm{g}$ and (d) the finger width \emph{w}, respectively. We note that the measured devices shown in Figure 2c and 2d are picked from another wafer W2. The fabrication process of W1 is the same as W2, except performing an oxygen plasma descum after the EBL development, in order to remove the residual resist, which results in lower phonon propagation loss and higher $Q_\textrm{i}$.
  }
  \label{}
\end{figure}

The experimental $Q_\textrm{i}$'s of SAW resonators on AlN-on-sapphire are also obtained to further confirm the advantages of thin film SAWs. The SAW chips are wire-bonded to the gold-plated cooper PCB and enclosed by the oxygen-free cooper sample-box. The sample box is thermally anchored to the mixed chamber of a diluted refrigerator(DR) and cooled to a base temperature of 10 mK for measuring the microwave reflection spectra with a vector network analyzer (VNA). The internal losses of the cryogenic SAW resonator mainly stem from propagation loss, phonon beam diffraction and leakage loss through mirrors \cite{delsing2019,bell1976surface,morgan2010surface}. Thereinto, phonon propagation loss is predominantly determined by the surface scattering of fabrication imperfections, surface roughness, and the intrinsic propagation loss of the substrate including viscous damping and scattering by the grain boundary and defects of piezoelectric films \cite{slobodnik1970microwave}. Whereas the other loss sources are dependent on geometry parameters of the resonator. The phonon beam diffraction \emph{Q} ($Q_\textrm{d}$) is proportional to $(\frac{w}{\lambda})^2$, where \emph{w} and $\lambda$ are the finger length and the wavelength of the SAW resonators, respectively. As expected, increasing \emph{w} mitigates phonon beam diffraction, resulting in higher $Q_\textrm{i}$. The reflection coefficient $\lvert\Gamma\rvert$ of the Bragg mirror near $f_0$ can be emulated by $\tanh(r_\textrm{s}N_\textrm{g})$, where $r_\textrm{s}$ denotes the reflectivity per electrode, and $N_\textrm{g}$ is the number of periods in each mirror. Intuitively, increasing $N_\textrm{g}$ leads to higher reflection coefficient, confining more phonon energy in the SAW resonator. In this letter, we mainly optimize the loss from phonon beam diffraction and leakage to improve $Q_\textrm{i}$ of SAW resonators on AlN films by systematically tuning the parameters of \emph{w}, $N_\textrm{g}$ and the cavity length \emph{L}.

Figure 2a shows the measured magnitude $\lvert S_\textrm{11} \rvert$ of one SAW resonator, labeled as S1, versus frequency at single-phonon powers of $P_\textrm{in} \simeq -141 $ dBm. The multiple evenly spaced SAW modes occur due to the constructive interference between the two mirrors. More than 25 SAW resonators with various design parameters are measured to extract the single-phonon $Q_\textrm{i}$. For accurately tracing the evolution of internal loss, the mean $\overline{Q_\textrm{i}}$ averaging the central 6 modes in each resonator and the maximum value $Q_\textrm{im}$ among them are both plotted. With increasing \emph{L}, the mean $\overline{Q_\textrm{i}}$ at the single-phonon level gradually increases and eventually saturates to $4.4 \times 10^4$ when \emph{L} exceeds 1 mm. Meanwhile the free spectral range (FSR) inversely decreases from 34.42 MHz to 3.03 MHz, in quantitative agreement with the formula description $\delta f/f_0 =1/(L/p+1/|r_\textrm{s}|)$, yielding $|r_\textrm{s}|$ = 0.013 also consistent with the FEM simulations. The resonance frequency of our SAW resonator reaches 5.66 GHz, higher than the previous values reported in literatures and providing a wider frequency range for coupling to superconducting qubits \cite{manenti2016surface,magnusson2015surface,shao2019phononic,andersson2021acoustic}. The highest $Q_\textrm{i}$ in our devices is obtained to be $5 \times 10^{4}$, hence the frequency-quality factor products (\emph{fQ}), assessing the isolation from thermal phonons, approaches as high as $2.81 \times 10^{14}$. Figure 2c shows $\overline{Q_\textrm{i}}$ and $Q_\textrm{im}$ versus $N_\textrm{g}$ with the identical \emph{L} = 1840.65 $\mu m$ and \emph{w} = 150 $\mu m$. The mean $\overline{Q_\textrm{i}}$ almost remains unchanged with increasing $N_\textrm{g}$ above 450, indicating $|\Gamma|$ approximately equals to unity and makes a negligible contribution to the internal loss of SAW resonators. The finger width \emph{w} dependent $\overline{Q_\textrm{i}}$ and $Q_\textrm{im}$, with \emph{L} = 1840.65 $\mu m$ and $N_\textrm{g} = 500$, is investigated showing a peak at \emph{w} = 200 $\mu m$ in Figure 2d. Further increasing the width of the finger length will induce more dielectric loss because of the increased participation ratio, which hints us the trade-off between acoustic diffraction and transducer dielectric loss should be considered for optimizing higher-\emph{Q} SAW resonators.

\begin{figure}[htbp]
  \centering
  \vspace{0.5cm}
  \includegraphics[width=\linewidth]{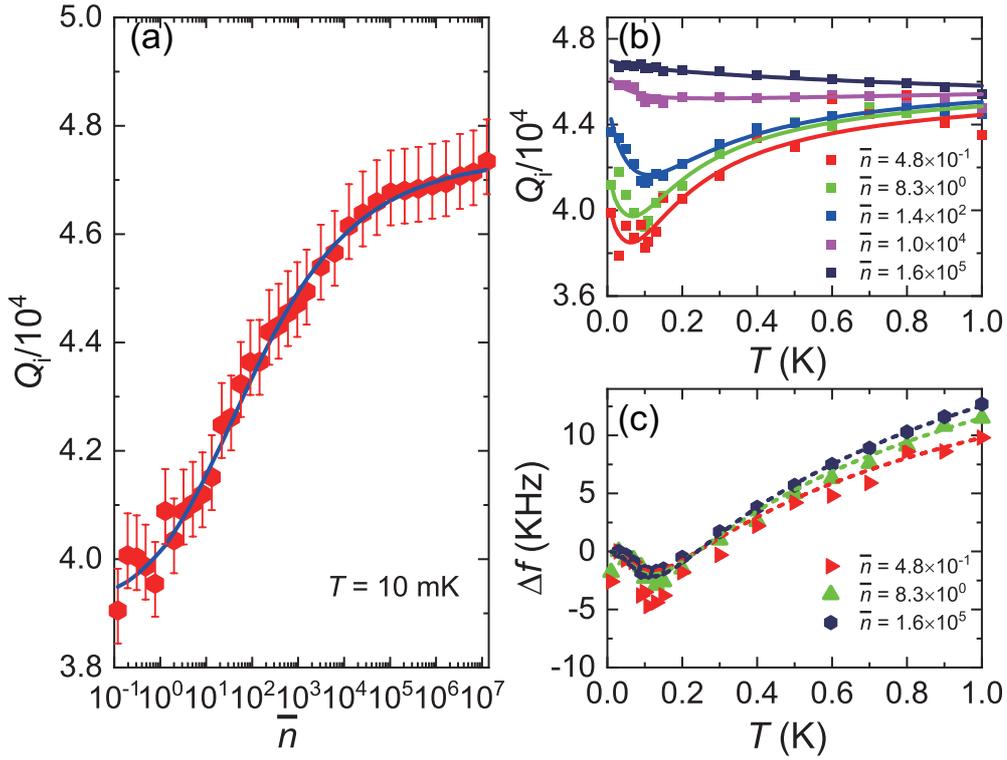}
  \caption{ Single-tone spectroscopy and analyses of phonon loss. (a) Phonon number dependence of the extracted $Q_\textrm{i}$ for the mode $f_0$ = 5.5976 GHz in the device S2 at 10 mK and the corresponding TLS model fitting (solid line). (b) Extracted $Q_\textrm{i}$ as a function of temperature at various phonon numbers $\bar{n}$. The solid lines show the fitting results based on the temperature dependent TLS model. (c) The frequency shift $\Delta f$ vs. temperature and the corresponding fitting (dash lines) using eq 2 for various phonon numbers.}
  \label{}
\end{figure}

Figure 3a shows the average phonon number dependent $Q_\textrm{i}$ for the SAW mode $f_\textrm{0}$ = 5.5976 GHz in another device S2. With increasing phonon numbers, $Q_\textrm{i}$ increases and eventually saturates to a constant value, which is also observed in SAW resonators on bulk piezoelectric substrates \cite{manenti2016surface, andersson2021acoustic}. This behavior is attributed to the saturation of two-level systems (TLS) at high powers which can be described with the standard tunneling model of TLS \cite{pappas2011two},
\begin{equation}\label{(1)}
\frac{1}{Q_\textrm{i}} =\frac{1}{Q_\textrm{TLS}} \frac{\tanh (\frac{hf_0}{2k_\textrm{B} T})}{\sqrt{1+(n/n_\textrm{c})^\beta}}+\frac{1}{Q_\textrm{rl}},
\end{equation}
where $Q_\textrm{TLS}$, $n_\textrm{c}$ and $\beta$ are the internal-\emph{Q} factor from TLS loss, the critical phonon number and the phonological parameter, and $Q_\textrm{rl}$ is the residual TLS-independent internal-\emph{Q} factor, including the aforementioned losses from phonon propagation, phonon beam diffraction and leakage. The fitted $Q_\textrm{TLS}$ and $Q_\textrm{rl}$ are $2.23 \times 10^5$ and $4.74 \times 10^4$, respectively. We fitted the power dependent $Q_\textrm{i}$'s for all the SAW modes as well, revealing about four times of $Q_\textrm{TLS}$ larger than $Q_\textrm{rl}$, which demonstrate that $Q_\textrm{i}$ is dominated by the residual TLS-independent loss. Because the losses from phonon diffraction and leakage are already optimized to negligible levels in our devices, we affirm that the dominant loss channel stems from phonon propagation at this stage.

\begin{figure}[h]
  \centering
  \vspace{0cm}
  \setlength{\belowdisplayskip}{0pt}
  \setlength{\abovedisplayskip}{0pt}
  \includegraphics[width=0.9\linewidth]{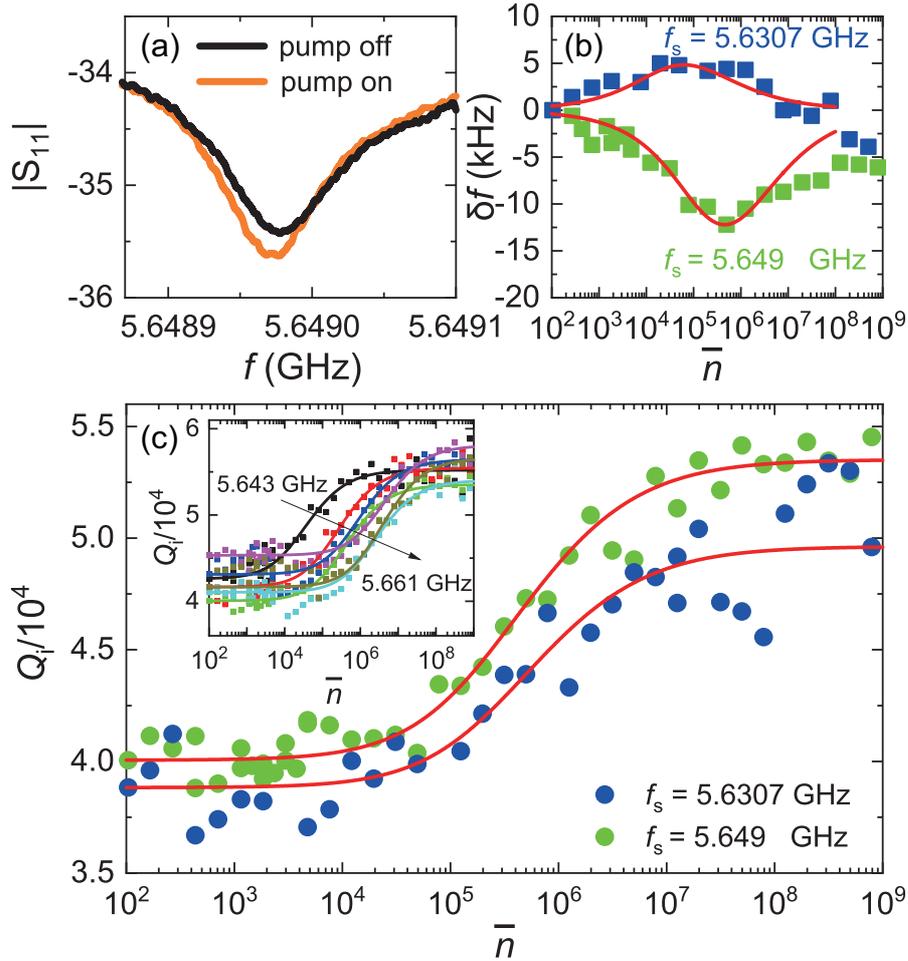}
  \caption{ Two-tone spectroscopy and analyses of the TLS ensemble. (a) The comparison in the reflection coefficient $|S_\textrm{11}|$ of one SAW mode $f_s = 5.649$ GHz with applying a strong detuned pump. (b) The measured frequency shift $\delta f$ of two probe modes with respect to the pump phonon number $\bar{n}$. The solid lines represent the fitting with the TLS model in the presence of the pump field. (c) Evolution of $Q_\textrm{i}$ with increasing $\bar{n}$ for two probe modes and the associated fittings. The inset plots $Q_\textrm{i}$ of the SAW modes, situated on the right side of the pump tone, as a function of pump phonon number.}
  \label{}
\end{figure}

The temperature dependent $Q_\textrm{i}$ and the resonance frequency shift $\delta f$ are obtained by measuring $S_\textrm{11}$ of the same SAW mode at temperatures ranging from 10 mK to 1 K, see Figure 3b and 3c. With increasing temperature, $Q_\textrm{i}$ starts to decrease below 100 mK and subsequently increases to a saturated value at 1 K, near $T_\textrm{c}$ of Al, which suggests that thermal quasiparticle-induced loss plays no role in SAW resonators. The TLS-related pronounced dip feature in $Q_\textrm{i}(T)$ has been confirmed by increasing driven powers which saturate the resonant TLS near the SAW mode, and bring about the vanishment of the dip. We believe that the abnormal decrease of $Q_\textrm{i}$ at low temperatures is caused by the spectra diffusion of TLSs, which has been recently observed in Tantalum CPW resonators \cite{dutta2022study}. $Q_\textrm{i}(T)$'s at various phonon numbers can be nicely fitted to the interacting TLS model accounting the temperature-dependent saturation term $n_\textrm{c}$ \cite{burnett2014evidence}. Moreover, the resonance frequency shift $\triangle f$ as a function of temperature is plotted and numerically fitted with the formula,
\begin{equation}\label{(2)}
\frac{\delta f}{f_0} = \frac{1}{\pi Q_\textrm{TLS}}\left\{\textrm{Re}[\Psi(\frac{1}{2}+\frac{hf_0}{2\pi i k_\textrm{B}T})]-\log (\frac{hf_0}{2\pi k_\textrm{B}T})\right\},
\end{equation}
with $\Psi$ representing the complex digamma function \cite{pappas2011two, mcrae2020materials}. The non-resonant TLSs entirely contribute to the frequency shift, also providing the evidence that the TLS loss is the major dissipation channel, except the phonon propagation loss in SAW resonators operating at mK temperatures.

To further extract the dephasing time $T_2$ of the TLS ensemble, we have performed
the two-tone measurements on the device S1. A drive tone with $f_\textrm{p}$ = 5.63984 GHz is applied to pump the center mode, while $S_\textrm{11}$ of other modes at weak probed power is monitored with sweeping pump power. The asymmetric saturations of off-resonant TLS give rise to the enhancement of $Q_\textrm{i}$ and the dispersive shifts of the resonance frequency derived from the Jaynes-Cummings Hamiltonian of the TLS coupled with the cavity mode, as displayed in Figure 4a \cite{andersson2021acoustic,kirsh2017revealing,capelle2020probing}. We elaborate the response of two adjacent SAW modes at 5.6307 GHz and 5.649 GHz, respectively. With increasing pump power, the positive (negative) detuned pump induces significant positive (negative) frequency shifts, in accordance with the TLS model in the presence of strong pump field \cite{kirsh2017revealing}. It's noteworthy that the discrepancy between $\delta f$ and the fitting at elevated powers is probably caused by high order nonlinearity of SAW resonators \cite{nakagawa2016discussion, tian2021research}. The strong off-resonant pumping field saturates TLS near $f_\textrm{s}$ and decreases the internal loss, as shown in Figure 4c. By independently fitting $\delta f(\bar{n})$ and $Q_\textrm{i}(\bar{n})$, We extract the average single-phonon Rabi frequency of the TLS ensemble being $\Omega_0 \approx$ 20-56 kHz, which is close to the reported value in SAW resonators on GaAs \cite{andersson2021acoustic}. Rewriting eq 1 with parameters of $\Omega_0$ and $T_2$, we obtain the average $T_\textrm{2}$ of the TLS ensemble is about 2.5 $\mu$s in SAW resonators on AlN-on-sapphire.

\section{Conclusion}\label{sec3}

In conclusion, we demonstrate SAW resonators on piezoelectric AlN films with the resonance frequency $f_0$ of about 5.66 GHz and the mean $\overline{Q_\textrm{i}}$ of $4.4 \times 10^4$, the highest $Q_\textrm{i}$ even reaching $5 \times 10^4$, in the quantum regime. The high coherence of SAW mechanical modes, along with the developed AlN-etching process lays the foundation for attaining the long-lived quantum state in SAW-qubit coupling quantum system. We have performed a comprehensive study on the internal loss of thin film SAW resonators through optimizing the geometry layout, alongwith the detailed temperature and power dependent $Q_\textrm{i}$ measurements. We deduce that the dominant loss of SAW modes originates from phonon propagation loss, secondly from TLS loss, and the possibilities of quasiparticle-induced loss are excluded in superconducting states. High-quality AlN films could further improve SAW resonators $Q_\textrm{i}$ with smoother surface and better crystallization. The TLS behavior and the dephasing time are investigated in detail, indicating the presence of significant TLS interactions in the phonon bath by the observed spectra diffusion. Finally, we stress that this heterogeneous approach, piezoelectric AlN thin films on low-loss sapphire substrates, provides an ideal material platform to experimentally implement quantum transducers converting the quantum state from microwave to optical domains, because of its fabrication ability, and high coherence of the mechanical resonator revealed in this work and the optical resonator \cite{liu2017aluminum, lu2018aluminum, fan2018superconducting}.

\section{Methods}\label{sec4}

The SAW resonator devices are designed with the coupling-of-modes (COM) model and fabricated on 400 nm of AlN films deposited on 420 $\mu$m thick sapphire with two-step lift-off methods. To pattern dense nanostructure of the transducer and Bragg mirrors in a SAW resonator, Bilayer electron beam resists, forming undercuts for the lift-off process, are coated on the surface and exposed using electron beam lithography (EBL). After developing, 50 nm of Al is deposited and subsequently lifted off. The ground plane and feedlines are formed by direct writing laserlithography and lift-off of the second Al layer. The AlN films are wet-etched in diluted H$_3$PO$_4$ to intentionally place superconducting qubits on low-loss sapphire. The microwave reflection spectra are measured at 10 mK with the vector network analyzer (Keysight N5231B). The input microwave signals are highly attenuated by the total of 76 dB distributed at each stage of the diluted refrigerator to suppress the thermal noise. The reflection signals routing from a circulator are sent to two isolators (LNF-ISC4\underline{~}8A, LNF-ISISC4\underline{~}8A), a commercial high electron mobility transistor (HEMT) amplifier mounted at the 4 K stage and the low noise amplifier at room temperature. Then, the output signals are acquired with the input port of VNA.

%

\backmatter

\bmhead{Acknowledgments}

We acknowledge the support from Superconducting Electronics Facility (SELF) in SIMIT for device fabrication. We thank Xin Ou, Yuandong Gu and Ya Cheng for insightful discussions. This work is partially supported by the Shanghai Technology Innovation Action Plan Integrated Circuit Technology Support Program (Grant No. 22DZ1100200), the National Natural Science Foundation of China (Grant No. 92065116), Strategic Priority Research Program of the Chinese Academy of Sciences (Grant No. XDA18000000), and the Key-Area Research and Development Program of Guangdong Province, China (Grant No. 2020B0303030002).

\bmhead{Author contributions}

Z.L. and W.J. conceived the experiments. J.C. and W.J. designed and fabricated the devices with the help from X.L., Z.N., and K.L. W.J. and J.C. carried out the measurements and analyzed the data. W.J., J.C. and Z.L. wrote the manuscript with input from all co-authors. Z.L. supervised the project.

\bmhead{Supplementary information}

Details of device fabrication (the SAW resonators, wet etching of AlN, and superconducting CPW resonators), the FEM simulation and geometry layout design of SAW resonators on AlN-on-sapphire, the circle fitting of complex $S_\textrm{11}$ data, and the analyses of the TLS ensemble in two-tone spectroscopy.

\end{document}